\documentclass[onecolumn,12pt]{IEEEtran}
\usepackage{amssymb,amsmath,epsfig,graphics,graphicx,enumerate,float,subcaption,verbatim,xspace, color,url}
\usepackage{multirow}
\usepackage{epstopdf}

\newcommand{\nop}[1]{}

\newcommand{\ie}{{i.e.}\xspace}
\newcommand{\eg}{{e.g.}\xspace}

\IEEEoverridecommandlockouts

\begin{document}
\title {Adversarial Learning:  A Critical Review\\ and Active Learning Study\thanks{This research is supported in part by a Cisco Systems URP gift and AFOSR DDDAS grant.}}
\author{\begin{tabular}{cccc}
 \small David J. Miller$^{1}$   &  \small Xinyi Hu $^{1}$  & \small Zhicong Qiu$^{1}$ & \small George Kesidis$^{1}$ \\
\small \tt djm25@psu.edu & \small \tt xzh106@psu.edu & \small \tt zzq101@psu.edu & \small \tt gik2@psu.edu\\
\end{tabular}\\ \vspace{3pt}
\small $^1$School of Electrical Engineering and Computer Science, Pennsylvania State University, University Park, PA 16802 USA\\
}
\maketitle
\thispagestyle{empty}
\pagestyle{plain}
\renewcommand{\baselinestretch}{1.50}

\begin{abstract}
This papers consists of two parts.  The first is a critical review of prior art on adversarial learning,
identifying some significant limitations of previous works.  The second part is an experimental study considering adversarial
active learning and an investigation of the efficacy of a mixed sample selection strategy for combating
an adversary who attempts to disrupt the classifier learning. 
\end{abstract}
\renewcommand{\baselinestretch}{1.30}

\vspace{-0.3cm}
\section{Introduction}
While there is still skepticism concerning the value of machine learning (ML) for network
security \cite{Paxson}, there has been growing interest in the ``dual'' problem
of investigating the {\it security} of 
ML systems, as applied both to {\it security-sensitive}
applications (network intrusion detection systems (NIDS), biometric authentication, email), as well as more generally (image,
character, and speech recognition, document classification), \eg
\cite{Tygar14,Tygar11,McDaniel16}.
Much of the focus is on
attempting to degrade or foil supervised classifiers, as well as anomaly detectors (ADs).  
\cite{Tygar11,Tygar14} provide a useful taxonomy for various attacks on classifiers, including whether they
affect training (what we will call {\it tampering}) or just testing/use ({\it foiling}).  Moreover, if the attack is on training, one
can distinguish attacks that involve {\it mislabeling}
from those that add {\it correctly} labeled examples, but ones with
contrived features chosen to bias learning.

\cite{Tygar11} demonstrated naive Bayes spam filters can easily be degraded by labeled spam examples (from known spam sources) 
which use many tokens
that commonly appear in normal (ham) email (an ``indiscriminate dictionary attack'').  This attack on training destroys discrimination power of most tokens in
the ``dictionary''.  Huge false positive rates ensued when as little as 5\% of training
data consisted of these contrived emails.  
\cite{Tygar14} considered active learning
(AL), a promising framework for security applications, as 
the classifier can adapt to track evolving threats and also because
oracle labeling may discover 
{\it novel} classes \cite{Qiu16}, \cite{ICASSP17} that may be zero-day threats. \cite{Tygar14} demonstrated, using SVMs, that 
if an adversary ``salts'' the unlabeled data batch in a {\it biased} fashion 
near the current decision boundary (where
AL seeks to choose samples for labeling),
one can induce classifier degradation -- each  
adversarial sample was chosen such that, if labeled, it will decrease accuracy the most.  
We will discuss in the sequel that the approach in \cite{Tygar14} appears to rely on (oracle) {\it mislabeling}, even though frequent mislabeling is not too realistic in practice.  We will also demonstrate
experimentally that such mislabeling is not necessary in order for the attacker to degrade classification accuracy.

\cite{McDaniel16} considers classifier testing/operation,
constructing examples that will be classified {\it differently} by a classifier than by a human being.  This
is an attack, \eg in the context of human and autonomous drivers, where one sees a STOP
sign and the other does not, or in future man-machine interactions where robot servants may misunderstand
their master's commands.  \cite{McDaniel16} showed that one could make {\it relatively} small perturbations 
to images of digits ({\it presumably} below human visual detectability, although we discuss this
further in the sequel) that alter a deep neural net's
decision.  Effectively, \cite{McDaniel16} minimally ``pushes'' patterns across 
the decision boundary.  While not recognized in \cite{McDaniel16}, their perturbation approach is related to boundary-finding algorithms
(neural network inversion) \cite{Hwang92,Hwang97}. 
Foiling has also been performed in other domains,
\eg against detecting 
malware in PDF documents: synthetic ``mimicry"   attacks
\cite{Smutz12,Laskov13} or 
natural documents with embedded malware 
\cite{Maiorca13} (``reverse" mimicry, as a trojan). 

In the related study \cite{Wagner16}, 
the attacker generates ``obfuscated'' voice commands
typically perceived as background noise 
by any human being who happens to be listening, but which
are recognized as valid commands by a speech recognition system.  
Such commands could be used \eg for financial fraud or to perpetrate a terrorist attack (controlling a crane, a train, etc.).
To defend against this,
they suggest, {\eg}:
a challenge (\eg CAPTCHA, reactive two-factor authentication); 
supervised (speaker-dependent) training to recognize only
authorized individuals; 
password\footnote{Recently, a child
used Amazon Echo to place an order - the parents had not set up
controls (a password) to prevent this \cite{Amazon-Echo}.},
speaker/voice authentication,
or some other kind of proactive two-factor authentication;
training a classifier to discriminate
between computer-generated obfuscated commands and nominal 
human commands;
increasing command specificity.  Also, disadvantages of these 
defenses are discussed, \eg latency and additional human effort associated with \eg a CAPTCHA 
for each utterance\footnote{This is completely
reasonable in some cases (\eg accessing bank accounts or entering
passwords), but may be considered very inconvenient in others
(\eg casual web surfing).}.

Regarding the latter defenses, the 
idea is to reduce the allowed variation of utterances classified
to each word. 
Thus, effectively, an ``unrecognized" (anomalous) class surrounds
each word in feature (\eg cepstral or wavelet coefficient) space.
Note that such AD in speech recognition is not new, \eg
\cite{Atlas90}, \cite{Borges08}.
Also, iPhone's Siri voice interface added ``individualized speech" 
(voice/speaker) recognition in 2015 \cite{Siri-speaker-recog}
(before \cite{Wagner16} but, again, this is very old technology, and
easily configured upon purchase of the phone).

Still, the threats posed by the attacks
like \cite{Wagner16}
need to be addressed\footnote{As do
other security concerns involving Siri, 
\eg \cite{Siri-passcode}, 
though it is not clear that one could use Siri to input data into web pages
via Safari on the iPhone.}. 
Supervised learning to discriminate the attack from ``valid" speech, which assumes either that {\it labeled} examples of the attack (and a sufficient {\it number} of these)
have been captured {\it or} that the attack strategy
is {\it known} (so that labeled examples can be synthesized) may not be realistic\footnote{The attacker may have great degrees of freedom he can apply to create inobtrusive examples -- it may be both difficult and impractical to try to create representative supervising examples ``covering'' all of these possibilities.}. 
Alternatively, we will advocate for an AD defense, which requires {\it neither} detailed knowledge of
the attack {\it nor} (up front) labeled examples of same.  

All of the above research works purport to demonstrate ``security holes'' in ML techniques.  These 
examples are provocative and they motivate further research.
However, these
studies also make convenient assumptions, \eg about the attacker's knowledge, which may be {\it grossly}
unrealistic.  Moreover, these studies ignore {\it existing} ML
techniques that are much more resilient to adversaries, even {\it without} considering {\it explicit} (and well known)
defenses, which may detect (and thus foil) the attack.  We next identify key limitations of some
prior adversarial learning works. 

\noindent
{\bf Unknown Classes and Authentication:}
\cite{Tygar11,Tygar14,McDaniel16}
assume each pattern {\it must} be classified to one of a {\it known} set of
categories.
In many systems,
in fact, there is an augmented class space with an ``unknown'' 
(unrecognized) category.
Patterns not confidently assigned to any known category may be
assigned ``unknown'' -- the attack examples in \cite{McDaniel16} could be 
assigned thus\footnote{For example, Siri responds ``don't know" to speech it
does not recognize as an english word, where every english
word (or word-tense combination) here corresponds to a known
category Siri has been trained to recognize.}.  
Moreover, in an AL setting, human oracles
who cannot confidently classify selected
samples may {\it reject} the sample (and thus the attack).
Likewise, emails that are part of an indiscriminate dictionary
attack could easily be detected as anomalous even relative to the existing ``spam'' class, could thus be labeled as ``unknown'', rather than ``spam'', and then would not be used
to help learn (\ie corrupt) the ``spam'' model.
Also,
particularly in IDS settings with {\it unknown unknowns}, the full complement of classes
is {\it a priori} unknown and one may {\it discover} new classes in an unlabeled or semisupervised data batch
\cite{Miller_Browning}.
This is especially true in an AL context, starting from few labeled examples
\cite{Qiu16}.
Finally,
many security applications in fact involve {\it authentication}, {\it not} classification
per se -- \cite{McDaniel16,Wagner16} assume the problem is classification.  The difference between these problems is well-known --
the latter {\em assumes}
that a datum
originates from one of the $N$ known classes
whereas
the former allows for the possibility that the datum is not associated
with a known class -- assignment to
the ``unknown unknown" class amounts to AD.
As an obvious example, consider a
perimeter authenticator (access controller) based
on a challenge for a userid and password;
here the
number of correct responses ($N$ authorized individuals)
is miniscule compared to the total number of possible inputs,
the vast majority of which result in failed access.
Moreover, authentication (at least to enter the perimeter) may require an {\it exact} (password) match for access.
This is essentially an extreme
example of giving emphasis to class specificity.  
Biometric based challenges (with well-known ``replay attack" protections)
can be used instead to explore
usability/security trade-offs, or to further secure access based on passwords.

\noindent
{\bf Unrealistic Assumptions:}
\cite{Tygar11}, \cite{McDaniel16}, and \cite{Wagner16} assume the classifier, both its structure 
{\it and its learned parameters}, are known to the attacker.  
\cite{Tygar14} further assumes that the {\it true}
joint (feature vector, class) distribution is {\it also} known to the attacker.
Knowledge of this distribution is usually the ``holy grail'' 
in ML (\ie it is {\em never} assumed known)
-- given it, one can form the Bayes-optimal classifier.
In practice, {\it at best}, one can only imperfectly
{\it estimate} this joint distribution, given a finite training set.  
So this latter assumption is wholly
unrealistic even for an inside attacker.  
Even ignoring this latter assumption, the former one -- 
full knowledge of the classifier --
is really only reasonable under two cases: 
1) a {\it non-security} setting, where public-domain
classifiers {\it might} be used;
2) in a ``security" setting,
if the attacker is in fact an {\it insider}, 
\eg if they work for the company that designed the classifier.  
Otherwise, in 
\eg biometric authentication, 
where one seeks to gain access to sensitive or restricted resources,
there is {\it zero} incentive to publicize many details of the 
authentication system.  \cite{Tygar14}
{\it further} assumes that the classifier is precisely known 
{\it after every round of AL oracle labeling}.  
Note that such knowledge might only be obtainable by intensive probing of the
system (to ``relearn'' the decision boundary) {\it in between} active labelings.  Such probing may be 
very resource-intensive and must complete within the limited time window between labelings.  Moreover, \cite{Tygar14}
considers just a 2-D example.
The number of probes needed to accurately learn the decision boundary
will grow with the feature dimensionality -- 
consider, {\eg} document or image domains,
where one may easily work with tens to hundreds of thousands of features.

\noindent
{\bf Asymmetry: ``Straw Man'' vs. A Robust System:}
While \cite{Tygar11,Tygar14} discuss possible defenses, 
the example attacks in 
\cite{Tygar11,Tygar14,McDaniel16} are on {\it completely
defenseless} systems, as well as {\it inherently} vulnerable ones.  
Even without explicitly
building defenses, there are ML techniques that are widely used and which
{\it also} (as a side benefit) make the system robust to exploits.  Consider the naive Bayes (NB) spam filter
attacks \cite{Tygar11}, including the ``indiscriminate dictionary'' attack
and the {\it red herring} attack, applicable both to email and to NIDS.  Here, spurious tokens are 
introduced into samples that come from a known spam source (\eg an email address recognized
as a source for spam).
Once the NB classifier ``takes the bait'' and adapts its classifier to focus on these
(apparently) discriminating tokens, subsequent spam emails (from various addresses) {\it omit} these tokens and thus avoid detection.
The reason such attacks are successful is because NB is a {\it weak} classifier, building only
a {\it single} model to represent a spam class that may (in a time-varying fashion) exhibit great diversity --
\ie NB effectively puts all its eggs in one basket.  Suppose, rather than a single NB model,
that one uses a {\it mixture} of NB models, with new mixture components introduced in a time-varying fashion,
as needed, to well-model patterns that are not well-fit by the existing model.  Such a mixture can capture and
{\it isolate}
a red herring attack within a single (new) NB component.  Thus, the main (legitimate) NB components representing the
spam class will not
be corrupted by the attack.  Moreover, once the attack is over, the probability {\it mass} of this
component will dwindle and this component can eventually be removed.  Thus, we
suggest a dynamic mixture, responsive to attacks (as well as time-varying classes), which should be highly
robust to red herring and indiscriminate dictionary attacks.  Other ML techniques that may help achieve
attack robustness include {\eg} ensemble 
classification  -- here, {\it uncompromised} classifiers may compensate for compromised ones,
helping to achieve robust
decisions via ensemble decision fusion.

The attacks in \cite{Tygar14,McDaniel16} 
could likely be defeated by relatively simple AD defenses.
For example, the AL attack in \cite{Tygar14} ``tricks'' the classifier
to select for labeling {\it biased}, attacker-generated samples near the current decision boundary.  
To maximize the success of the attack,
these synthetic examples should be chosen for active labeling 
with {\it high prevalence}.
Moreover, as elaborated in the next section, this attack appears to rely on the assumption that the oracle
may {\it mislabel} samples near the true (optimal) decision boundary.
As demonstrated in the next section: 1) a successful attack does not require oracle mislabeling;
2) irrespective of whether there is such mislabeling, attacks can be defeated
by using a {\it mixed} strategy for AL sample selection (not always selecting the sample nearest
the current decision boundary), and with no significant loss in the efficiency and accuracy 
of classifier learning.

Likewise, \cite{McDaniel16} effectively assumes that if the {\it number} of altered features (image pixels, for character recognition)
is below a fixed threshold, the attack will not be detectable by a human being.
First, this is questionable,
as the resulting salt and pepper noise is 
{\it quite} visible (see Fig. 1 in \cite{McDaniel16}) and is {\it not} typical of the original (clean) images in the database\footnote{Human subject testing
{\em was} used in \cite{McDaniel16}.  However, the authors did {\it not} ask respondents whether they thought images
had been tampered with --they only asked them to classify the images.}. 
Second, whether or {\it not} the attack is perceptible to a human, it may be {\it easily} detected by an AD
defense.  Peculiarly, \cite{McDaniel16} limits the number of features (pixels) one is allowed
to alter.  In so doing, to induce a classification error, the magnitudes of the perturbations of the chosen features must be substantial\footnote{Even though the authors impose the {\it minimum} norm perturbations needed
to induce classification errors (note again that the introduced salt and pepper noise is quite
visible).}.  Accordingly, an AD may easily detect salt and pepper noise pixels as anomalous, relative to
intensity values of pixels in a surrounding local spatial neighborhood.  
Finally, we note that if a human and machine {\it do} disagree on an example -- but if they can
{\it share} their decisions -- then the introduction of these ``attack'' examples becomes an {\it opportunity} -- to actively learn -- so as
to rectify the machine's decision on such examples\footnote{This assumes that the human is more accurate than the machine.  In some application domains, this is certainly the case.  In domains where this is not the case,
such disagreement may by the same token (appropriately) cause the human to reconsider their judgement.}.

\noindent
{\bf Tampering Power:}
~In \cite{Tygar11}, 
while the authors {\it limit} the power of the attacker
to corrupt training data (5\% of email training data),
it is unclear that this level of tampering power is plausible/corresponds to a realistic scenario. 
In principle, if the
attacker is an insider, she may have unlimited capability to corrupt training data, in which case there may
be no adequate defense.  Otherwise, the chosen 
power of the attack 
may intricately depend on knowledge of the particular defense system
in play, \ie on the attacker's {\it tradeoff} between achieving high attack potency and minimizing the probability
of the attack's detection and thwarting.  
Even if only a few training samples are altered, this tradeoff may exist.  For example, 
in \cite{Xiao15}, tampering with a single support vector is show to dramatically degrade
classification accuracy.  However, to achieve such effect, the tampered sample may become an extreme outlier of its class, and thus may either be ignored (via use of margin slackness)
or may be detectable as a suspicious sample. 

\noindent
{\bf Reverse Engineering:}
\cite{McDaniel16} (strongly) assumed that the classifier structure and its parameter values are known to the attacker.
Recent works \cite{Reiter16}, \cite{PapernotMGJCS16} have proposed techniques to reverse-engineer
a (black box) classifier without necessarily even knowing its structure.  In \cite{Reiter16}, the 
authors consider black box machine learning {\it services}, offered by companies such as
Google, where, for a given (presumably big data, big model) domain, a user pays for class
decisions on individual
samples (queries) submitted to the ML service.  \cite{Reiter16} demonstrates that, with a {\it relatively} modest number of queries (perhaps as many as ten thousand or more), one can {\it learn}
a classifier on the given domain that closely mimics the black box ML service decisions.  Once
the black box has been reverse-engineered, the attacker need no longer subscribe to the
ML service.  One weakness of \cite{Reiter16} is that it neither considers very large (feature
space) classification domains nor very large networks (deep neural networks (DNNs)) -- {\it orders}
of magnitude more queries may be needed to reverse-engineer a DNN on a large-scale domain.  However, a much more critical weakness of of \cite{Reiter16} stems from one of its (purported) greatest
{\it advantages} -- the authors tout that their reverse-engineering does not require {\it any} labeled training
samples from the domain\footnote{For certain sensitive domains, or ones where obtaining real examples is expensive, the user may in fact have no realistic means of obtaining a significant number of real data examples from the domain.  This is one main reason why the ML service is needed in the first place -- the company or its client for this domain are the (exclusive) owners of this (labeled, precious) data resource.}.  In fact, in \cite{Reiter16}, the attacker's queries to the black box
are {\it randomly} drawn, \eg uniformly, over the given feature space.  While such random querying 
is demonstrated to achieve reverse-engineering, what was not recognized in \cite{Reiter16} is that
this random queryinig makes the attack {\it easily detectable} by the ML service -- randomly
selected query patterns will typically look nothing like legitimate examples from any of the classes --
they are very likely to be extreme outliers, of all the classes.  Each such query is thus
{\it individually} highly suspicious by itself -- thus, even tens, let alone thousands of such queries
will be trivially detected as jointly improbable under a null distribution (estimable from the training set defined over all the classes from the domain).  Even if the attacker employed bots, each of
which makes a small number of queries (even as few as ten), each bot's random queries should be easily detected
as anomalous, likely associated with a reverse-engineering attack.

We next perform an experimental study involving adversarial active learning.
\vspace{-0.3cm}  
\section{Active Learning Experimental Setup}
In \cite{Tygar14}, as noted earlier, the attacker can perfectly 
estimate the current decision rule after every AL round (may require lots of probing between rounds) and knows the true joint density on the feature vector and class label $p(Y,X)$ (wholly unrealistic). It was furthered assumed that the attacker knows the AL sample selection strategy (uncertainty sampling) and injects one new (high decision uncertainty) sample 
into the unlabeled batch at each AL round, crafted so that it will be chosen by the
oracle for labeling.
Moreover, from the description given in \cite{Tygar14}, the authors do {\it not} assume that the oracle
labels the attacker's sample consistent {\it either} with the Bayes-optimal decison rule {\it or} randomly, according
to the true class posteriors.  That is, in \cite{Tygar14}, although it is not very clearly stated, the oracle may
{\it mislabel} the samples crafted by the attacker.  This is crucial to the success of the attack in \cite{Tygar14}. 

In this paper, we propose a more realistic framework, under which the attacker still possesses the ability to degrade
classification accuracy even without the unrealistic oracle mislabeling assumption.  However, we also demonstrate that
attacks on AL can be defeated by a {\it mixed} sample selection strategy, through which the attacker's injected samples
are not {\it so} frequently chosen for oracle labeling.  Moreover, this mixed strategy
does not make a significant sacrifice either in classifier accuracy or learning convergence (number of queries needed to achieve good accuracy). In fact, this strategy is also suitable for defeating the attack
even in the presence of oracle mislabeling (though we focus on more realistic oracle labeling in the sequel).
\vspace{-0.3cm} 
\subsection{Preliminaries}
\label{sec:setup}
As in \cite{Tygar14}, we assume a two-class problem and use a two-class linear support vector machine. We first select a (large) (unlabeled) training pool ($T_r$), then randomly select a relatively small number of samples from both classes in $T_r$ and assign (ground-truth) class labels to them\footnote{For our synthetic data experiment, consistent with the ground-truth class distributions from \cite{Tygar14}, these labels are assigned according to the Bayes-optimal rule.  For our real-world digits experiment, we use the labels provided with the given data set.}. 
One half of this labeled subset is used as a labeled training set ($T_l$) to train the initial SVM classifier. 
The other half of this labeled subset is used as a ``validation'' set ($V$), to estimate an approximate class posterior for the SVM, using the approach described in \cite{Platt}. The (large) remainder in $T_r$ is taken as the pool of unlabeled samples available to the active learner ($T_u$), from which the active learner selects for Oracle labeling (and into which the attacker inserts adversarial samples). In addition, there is a labeled test set to evaluate classifier accuracy.
\vspace{-0.3cm} 
\subsection{Sample Selection Criteria for Active Learning}
\label{sec:setupactive}
The active learner selects samples from $T_u$, one by one, for labeling by the oracle, with the SVM classifier retrained after each oracle labeling. We investigate the following AL sample selection strategies:

\noindent
{\bf Uncertainty sampling:} Uncertainty sampling is the simplest and most commonly used sample selection strategy \cite{Burr}. In this strategy, the AL chooses the nearest sample in $T_u$ to the current boundary (the most uncertain sample).

\noindent
{\bf Max-Expected-Utility (MEU):} Sampling to maximize expected utility (classifier gain) \cite{Hospe}. For data $x_i\in T_u$, $i^*={\rm{argmax}}_i \tilde{U_i}(\theta)$, where 

$\tilde{U_i}(\theta)=$ 
\begin{small}
\begin{equation}
\sum_{y_i}p_\theta (y_i|x_i)\frac{1}{N}\left( \sum_{j\in L\cup i} p_{\theta_{+i}}(y_j|x_j)+ \sum_{j\in U \setminus i} \sum_{y_j} p_\theta (y_j|x_j)p_{\theta_{+i}}(y_j|x_j)\right).
\label{equ:exputi}
\end{equation}
\end{small}
Here, $\theta$ is the current set of class posterior parameters, and $\theta_{+i}$ reflects the updated parameters after adding $x_i$ to $T_l$ with its {\it putative} label $y_i$. Using Platt probabilistic outputs for SVMs \cite{Platt}, the posterior probability $p_\theta (y_i|x_i)$ is based on a logistic regression approximation to
the SVM (hard) decision function. Finally, $N=|T_l|+ |T_u|$.

\noindent
{\bf Random sampling:} Select from $T_u$ according to a uniform distribution. This sample selection acts as a baseline. 

\noindent
{\bf Mixed strategies:} Choose the sample by MEU (or random sampling) with probability $p$; otherwise by uncertainty sampling with probability $1-p$.  Note that a non-zero proportion for uncertainty sampling is warranted because 
uncertainty sampling is a very good mechanism for discovering unknown classes that may be latently present in $T_u$ \cite{Qiu16}.
At the same time, using uncertainty sampling plays into the hands of the attacker. 
\vspace{-0.3cm} 
\subsection{The Attacker}
\label{sec:setupattacker}
We assume the attacker has the same knowledge as the active learner (\ie the attacker knows $T_l$, $T_u$, and the current SVM classifier boundary).
The attacker adds one sample to $T_u$ at each active sample selection round, chosen as follows:
1) he {\it projects} all training pool samples ($T_u$ and $T_l$) onto the current decision boundary (hence creating
a (rich) candidate pool of high uncertainty samples), as shown in the figure below; 2) the attacker injects into $T_u$ the candidate from this pool with {\it minimum} expected
utility, based on the expected utility objective given above.

\begin{figure}[htbp]
\begin{center}
\includegraphics[scale=0.4]{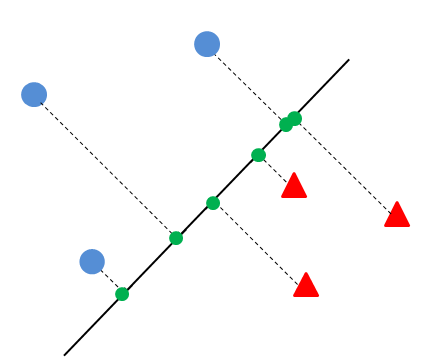}
\caption{Candidate adversarial samples (green dots on the boundary)}
\label{proj}
\end{center}
\end{figure}

\vspace{-0.3cm} 
\section{Synthetic Data Experiments}
\vspace{-0.3cm} 
\subsection{Dataset}
We first consider the two-dimensional feature space problem from \cite{Tygar14}. The data $X$ is generated such that 
the class 1 instances have a bivariate normal distribution centered at (2,0), with the class 2 instances bivariate normal centered at (-2,0); both classes use an isotropic (identity) covariance matrix, and the classes are equally likely. We generated 105 instances from each of the two classes as $T_r$ initially, from which we drew 5 samples at random from each class to form $T_l$, and 
another 5 samples from each class to form $V$.
The remaining 190 instances were taken as $T_u$. We also generated 200 instances from each class as test set. This dataset is (obviously) not linearly separable. The optimal decision boundary is the $Y$ axis. 
Our oracle deterministically assigns labels consistent with this optimal decision boundary.
\vspace{-0.3cm} 
\subsection{Results}
\begin{center} 
\begin{figure}[htbp]
\begin{minipage}{0.4\linewidth}
\includegraphics[scale=0.35]{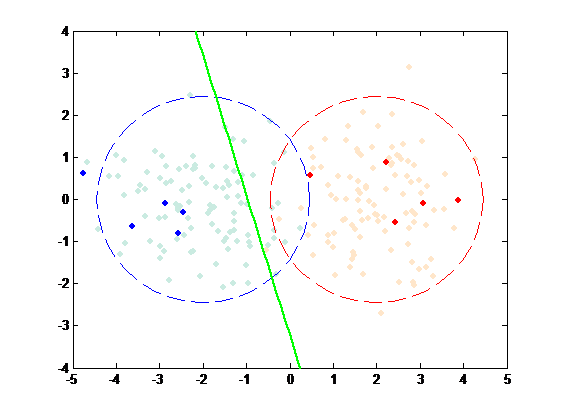}
\centerline{(a) initial sampling}
\end{minipage}
\hfill
\begin{minipage}{0.5\linewidth}
\includegraphics[scale=0.35]{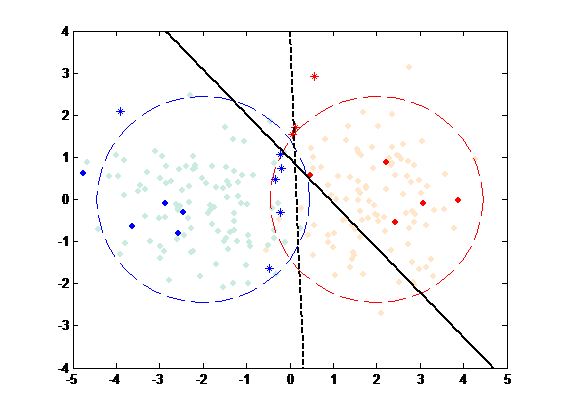}
\centerline{(b) after 9 queries}
\end{minipage}
\caption{Classifier decision boundary after initial training and after 9 queries without attack (dash lines) and with attack (solid lines).}
\label{fig:syn}
\end{figure}
\end{center}

First, we conducted a single experimental trial, for which AL selects samples strictly using uncertainty sampling;
thus, the attacker's adversarial samples are selected and labeled at each round\footnote{Note, though, that, unlike 
\cite{Tygar14}, the oracle does not perform any mislabeling.}. Fig. \ref{fig:syn} shows that the labeled adversarial samples induce a decision function that deviates from the optimal rule (the boundary becomes tilted from vertical). Hence, the attacker does have the ability to degrade AL classification accuracy (even without any oracle mislabeling). 


In the following experiment, we performed 10 random trials. In each trial, we used the same $T_r$, but randomly chose the initial $T_l$ and $V$ from $T_r$.  We computed average performance for different AL strategies, over these 10 trials.

\begin{figure}[htbp]
\begin{center}
\includegraphics[scale=0.48]{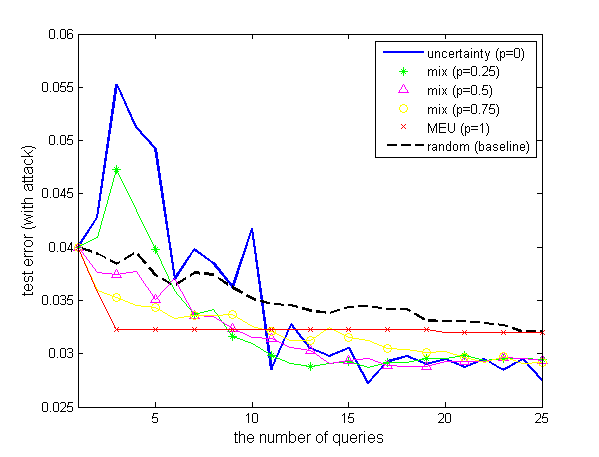}
\caption{Average test error performance of different strategies with attack on the synthetic dataset.}
\label{fig:syntest}
\end{center}
\end{figure}

\begin{figure}[htbp]
\begin{center}
\includegraphics[scale=0.48]{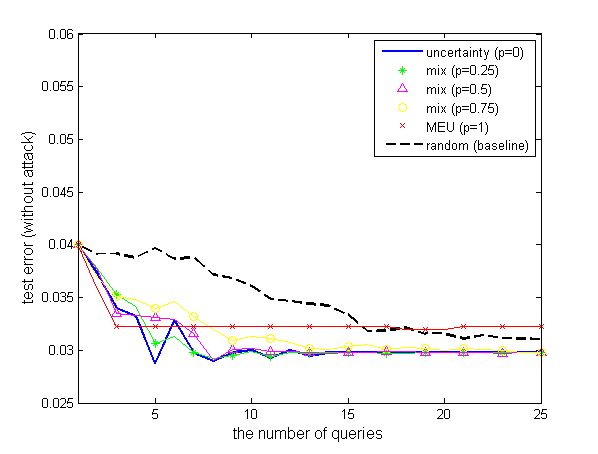}
\caption{Average test error performance of different strategies without attack on the synthetic dataset.}
\label{fig:syntestno}
\end{center}
\end{figure}
Fig. \ref{fig:syntest} and Fig. \ref{fig:syntestno} show performance both in the presence of and in the absence of the attack, respectively.
Different sample selection strategies show different abilities to defeat the attack, as shown in Fig. \ref{fig:syntest}.
We have the following observations:
\begin{itemize}
\item When AL uses strict uncertainty sampling $(p=0)$, adversarial samples degrade classification accuracy successfully for the first 15 queries, which is consistent with single-trial results. However, the fabricated samples the attacker inserts do not degrade performance in the long run.
Thus, the attack effectively 
only delays convergence to a good decision boundary\footnote{Mislabeling would likely allow {\it perpetuated} accuracy degradation.}. When the current boundary is very close to the optimal one, if the attacker still adds fabricated samples to the current boundary, these samples may even be helpful to refine the boundary.  As shown in the sequel, however, the attack is more successful on a real-world, high-dimensional digit recognition domain.
\item  Fig. \ref{fig:syntest} indicates MEU $(p=1)$ is not substantially affected by adversarial samples (as one would expect, since the MEU sample selection criterion is the antithesis of the adversary's sample generation strategy), and makes the test error decrease the most
at the beginning. However, we also noticed that MEU selects samples 
in a class-biased fashion (many samples from one class),
leading to a biased active learner after multiple queries. It is also clear from Fig.\ref{fig:syntestno} and Fig. \ref{fig:syntest} that MEU converges to a suboptimal decision boundary in the long run. 
\item With the attack present, the mixed strategies shows improved accuracy for increasing $p$, $p \in [0.25,0.75]$.  Moreover, in the absence
of the attack, there is little accuracy difference for different choices of $p$.
\item Note also that the random strategy is robust to the attack, but does not converge to a solution as accurate as the mixed strategies.
\end{itemize} 
\vspace{-0.3cm} 
\section{Handwritten Digit Experiments}
\vspace{-0.3cm} 
\subsection{Dataset}
We used the MNIST dataset, consisting of $28\times28$ pixel grayscale images, learning a linear SVM to
discriminate between the digits ``5" and  ``6", that is, we have 784 (pixel) features and a two-class problem.
We initially chose 105 ``5" digits and 105 ``6" digits as $T_r$ and again labeled 5 samples from each class to form $T_l$, and 5 samples from each class to form $V$. The remainder of $T_r$ was taken as $T_u$. Also, we randomly chose another (distinct set of) 456 ``5" and 462 ``6" samples to form a test set.
Our oracle is assumed to be 
an SVM trained on the entire data set.  Since the entire data set is linearly separable, this is a plausible choice
for the oracle.  Note, also, that because the data set is linearly separable, this oracle assigns {\it ground-truth} labels to all original data samples that
are chosen for oracle labeling -- the oracle only manufactures labels for the adversarial samples that are
selected for labeling (and in this case it does so objectively, consistent with maximum margin linear separation of the entire data set).  
\vspace{-0.3cm} 
\subsection{Results}
\begin{figure}[htbp]
\begin{center}
\includegraphics[scale=0.5]{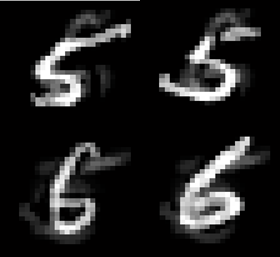}
\caption{Digit projections on the boundary, Oracle labeling (from top to bottom, from left to right): 6, 6, 5, 6}
\label{fig:digitproj}
\end{center}
\end{figure}

As described in \ref{sec:setupattacker}, candidate adversarial samples are the projections of all samples in $T_l$ and $T_u$ onto the current boundary. Some candidates shown in Fig. \ref{fig:digitproj} involve superposition of the two digits -- labeling such samples and subsequent classifier retraining is expected to have a (negative) impact on
the classifier decision boundary. 

\begin{figure}[htbp]
\begin{center}
\includegraphics[scale=0.48]{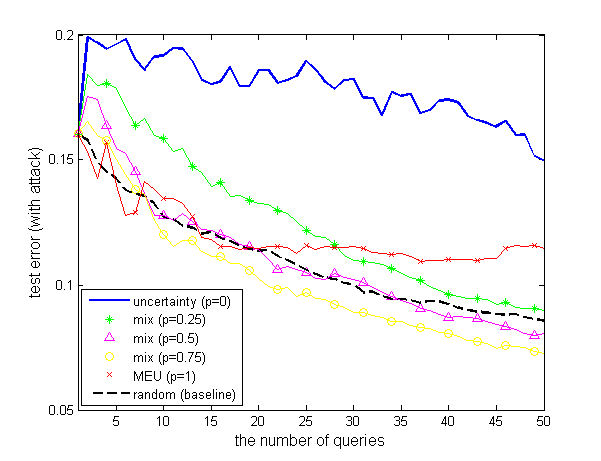}
\caption{Average test error performance of different strategies with attack on Digit Dataset.}
\label{fig:digittest}
\end{center}
\end{figure}

\begin{figure}[htbp]
\begin{center}
\includegraphics[scale=0.48]{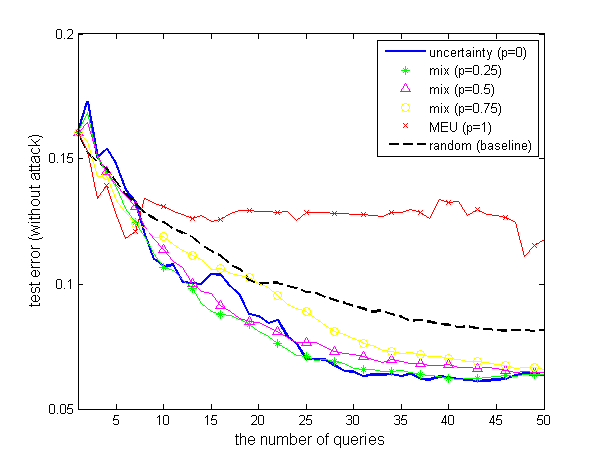}
\caption{Average test error performance of different strategies without attack on Digit Dataset.}
\label{fig:digittestno}
\end{center}
\end{figure}

Again, we performed 10 random trials to get the average performance with/without attack in Fig. \ref{fig:digittest} and Fig. \ref{fig:digittestno}, respectively.

To summarize the results:

\begin{itemize}
\item Using uncertainty sampling $(p=0)$, AL always selects the attacker's injected samples, and is the best strategy without attack, but worst with attack.  Note also that the attack much more substantially delays learning progress for this high-dimensional domain, compared with the 2-D example.
\item MEU $(p=1)$ makes the test error decrease the most at the beginning, but it converges to a suboptimal decision boundary, both with and
without the attack.
\item The mixed strategies have a good defensive capability against the attack (by sometimes using MEU), and they also alleviate the unbalanced selection problem of MEU by sometimes using uncertainty sampling. With the attack, there is monotonically improving performance with $p$, for $p \in [0.25,0.75]$ -- using the mixed strategy with $p=0.75$, Fig.  \ref{fig:digittest} shows the test error has a fast and steady decrease.  All the mixed strategies perform similarly without the attack.
\item The random strategy fares well with the attack, but not when the attack is absent. 
\end{itemize}
\vspace{-0.3cm} 
\section{Future Work}
In future work, we will investigate some of the other adversarial learning defenses suggested in our review section.  We may also investigate alternative AL sample selection strategies -- {\eg}, a modification of the MEU strategy that
does not suffer from the biased sampling we observed in our experiments (This may be achieved by estimating class proportions and modifying the MEU strategy to maximize expected utility, but while also sampling consistently with these class
prior estimates.). Further, we will continue to study mixed strategies to discover the unknown classes.

\bibliographystyle{plain}
\bibliography{../../latex/anomaly-detection,../../latex/kesidis-prior,../../latex/spam,../../latex/botnet,../../latex/references_Fatih,../../latex/MyCollection,../../latex/dns,../../latex/stavrou-prior,../../latex/active-learning,../../latex/transductive,../../latex/reputs,../../latex/adversarial,../../latex/speech,../../latex/ref-miller,../../latex/miller,../../latex/IoT}
\pagebreak

\end{document}